\documentclass[intlimits,twoside,a4paper]{article}

\usepackage[cp1251]{inputenc}

\usepackage[eqsecnum]{cmpj3}

\usepackage{bm}

\issue{2022}{25}{1}{13502}
\doinumber{10.5488/CMP.25.13502}

\title[Non-extensive thermodynamics of the radiation in heterogeneous thermal plasmas]%
{Non-extensive thermodynamics of the radiation in heterogeneous thermal plasmas%
}
	
\author[G. S. Dragan, V. V. Kutarov, A. Y. Bekshaev]
{G. S. Dragan\orcid{0000-0002-3626-9760}\thanks{Corresponding author: \email{dragan@onu.edu.ua}.},
        V. V. Kutarov\orcid{0000-0002-3236-6096},
        A. Y. Bekshaev\orcid{0000-0003-4153-559X}}
    
\address{Physics Research Institute, Odesa I. I. Mechnikov National University, Dvorianska, 2, \\Odesa, Ukraine, 65082
}

\date{Received September 10, 2021, in final form January 4, 2022}

\begin{document}

\maketitle

\begin{abstract}
Thermodynamic characteristics of the radiation of condensed combustion products presented in the form of agglomerates of metal-oxide nanoparticles demonstrate deviations from the classical Planck’s law. We propose to interpret these deviations in terms of the non-additive entropy of the photon system interacting with the heterogeneous combustion products, which makes it possible to use the non-extensive Tsallis thermodynamics for their description. It is assumed that the non-additive character of the radiation entropy in heterogeneous plasma can be explained by the influence of long-range interactions and non-equilibrium physicochemical processes. An expression is obtained for the energy-dependent distribution of the photon density, based on the phenomenological parameter of non-extensiveness $q$ which, in the first approximation, does not depend on the energy. In this case, the ''non-extensive'' Planck’s law can be reduced to the ``usual'' Planck distribution by introducing the ``effective temperature'' that exceeds the real temperature. Numerical modelling has shown that the spectral density of photons, the position and magnitude of its maximum depend on the value of the parameter~$q$, which can be used for its experimental determination and revelation of its physical nature and origin.	
%
%
\printkeywords
%
\end{abstract}

\section{Introduction}\label{sec1}


The radiation heat transfer plays a decisive role in a number of thermo-physical phenomena, from evolution of Universe to everyday life. In particular, this transfer is important in physics of technical plasma, including combustion of air-suspensions of metal powders \cite{Zolotko} and metallized compositions \cite{Vishnyakov}, in numerous industrial \cite{Lautenberger, Ning} and natural \cite{Lubin, Trigger} processes. Usually, the spectral composition of the radiation of nanoparticles and their agglomerates is well described using the equilibrium Planck distribution \cite{Planck}  with empirical values of the emissivity coefficient.

However, significant violations of the Planck law were  recently revealed for cosmic radiation \cite{Trigger, Zagorodny}, which was explained by the influence of plasma matter, special features of the density and of the motion of charged particles (grains). Analogous deviations from the Planck law were also observed in the combustion products of metallized fuel compositions \cite{Dragan}. Despite the huge difference in the spatial dimensions, both cases are united by the presence of interactions between  separate parts of the system (individual emitters): in astrophysics, this is due to their very large scale and immense mass obeying the long-range gravitational interaction, while in plasma, quite oppositely, this is due to microscopic grains intensively interacting via the Coulomb and atomic forces. In both situations, the main prerequisite of the classical thermodynamics is destroyed: the thermodynamic functions, first of all, the energy and entropy of the system, are not additive quantities, and the interaction of individual emitters is important.

For such systems of interacting radiators, approaches of nonextensive thermodynamics are developed and investigated \cite{Tsallis, Kolesnichenko}. All of them are aimed at removing or, at least, slackening one of the formal postulates of classical thermodynamics: the a priori assumption of smallness of the grain-interaction radius as compared with the overall dimensions of the considered thermodynamic system. This assumption corresponds to the principle of molecular chaos and enables the additivity (extensiveness) of  thermodynamic quantities, such as free energy, entropy, etc. However, this principle is violated in numerous examples of non-extensive thermodynamic systems, whose sizes are comparable with the characteristic radius of interaction between the grains that form them. The behavior of such systems cannot be represented by the Boltzmann-Gibbs statistics, and the statistical distribution functions become non-Gaussian. In particular, an adequate description of the thermodynamic properties of such “small systems” requires the higher-order statistical moments to be taken into account \cite{Hill}.

The reasons why the Boltzmann-Gibbs statistics fails to provide an appropriate description for a thermodynamic system are different. These can be, for example, “memory effects” that arise when its state depends on the parameters at previous times and violate the hypothesis of molecular chaos. Another example is supplied by the systems with long-range interactions of the plasma type, which should be considered as nonequilibrium (quasi-equilibrium) stationary states. Besides, the non-extensiveness of states in plasma may emerge from the inhomogeneity and anisotropy of the thermodynamic characteristics~\cite{Balescu}. 

Despite  different physical origins of  non-extensiveness, the corresponding effects can be formally analyzed with the help of general approaches employing the phenomenological generalization of the main thermodynamic relations to the non-additive conditions. One of the most promising methods was proposed by Tsallis \cite{Tsallis} where the entropy of the system of interacting radiators is represented by a non-additive function that depends on a certain control parameter $q$, and when the value $q \rightarrow 1$, the expression for the entropy reduces to its classical form. Remarkably, the Tsallis’s approach has shown its applicability to systems that are not in the equilibrium (although stationary) state: e.g., in the space applications \cite{Pavlos,Yoon2019,Yoon2020} as well as in general statistics and information aspects \cite{Strzalka,Ghanbari}.

In this article, based on the Tsallis idea, we apply the concepts of non-extensive thermodynamics to the radiation produced by a heterogeneous plasma with long-range interactions destroying the basic postulate of the molecular chaos. Such systems supply classical examples of the thermodynamic non-additivity; additionally, they are, in general, far from  equilibrium conditions and, strictly speaking, must be described by exquisite and complicated instruments of  non-equilibrium thermodynamics \cite{Groot,Olemskoi}. However, as we will show later on, a simplified phenomenological description employing the Tsallis’s $q$-parameter appears to be helpful in the analysis of their important features, in particular for the description of their radiation. Keeping in mind the specific example of metal-oxide nanoparticles agglomerates occurring in the combustion plasma, the approach developed is largely independent of the particular nature of the microscopic emitters and probably will be useful in a wider field of applications.

\section{The general approach} \label{sec2}

In the framework of the Boltzmann-Gibbs statistical thermodynamics, the entropy of a system of photons and photon emitters is determined by the expression \cite{Goldin}

\begin{equation} \label{eq1}
\begin{gathered}
S_i=k_{\text{B}}\ln W_i,
\end{gathered}
\end{equation}

\noindent where $W_i$ is the statistical weight of the microstate of the thermodynamic system, $k_{\text{B}}$ is the Boltzmann constant. The microstate is set, as a rule, using the photon energy distribution function. The implementation of macrostates is defined as the combination of a large number of microstates of the thermodynamic system represented, for example, in the form of an agglomerate of metal oxide nanoparticles.

We start with considering a usual approach to the thermodynamic system as the combination of many disjoint events. Then, the statistical weight of the macrostate of such a system $W$, consisting of some set of statistically independent subsets, can be determined as

\begin{equation} \label{eq2}
\begin{gathered}
W=\prod\limits_iW_i
\end{gathered}
\end{equation}

\noindent and the value of its entropy is additive (extensive):

\begin{equation} \label{eq3}
\begin{gathered}
S=\sum\limits_iS_i.
\end{gathered}
\end{equation}

\noindent The representation of the thermodynamic system entropy in the form (\ref{eq3}) is valid only in the approximation based on the hypothesis of the molecular chaos.

Let a system of emitters consist of nanoparticles interdependent via radiation. Then, the thermodynamic scope can be extended to systems with nonextensive properties of entropy as it was undertaken by Tsallis who axiomatically  introduced an expression for non-extensive (NE) entropy in the form \cite{Tsallis}

\begin{equation} \label{eq4}
\begin{gathered}
S_{\text{NE}}=k_{\text{B}}\frac{W^{1-q}-1}{1-q},
\end{gathered}
\end{equation}

\noindent where $q$ is a phenomenological parameter of non-extensiveness. For a thermodynamic system that is described by the Boltzmann-Gibbs statistics, the parameter $q \rightarrow 1$, and expression (\ref{eq4}) transform into the classical formula for entropy

\begin{equation} \label{eq5}
\begin{gathered}
S=k_{\text{B}}\ln W.
\end{gathered}
\end{equation}

The statistical weight of the photon system is defined as the number of ways by which a given macrostate can be realized. The number of ways of placing $n_i$ photons in $\textit{g}_i$ states is determined by the expression \cite{Goldin}

\begin{equation} \label{eq6}
\begin{gathered}
W=\prod\limits_i\frac{(n_i+\textit{g}_i-1)!}{n_i!(\textit{g}_i-1)!}.
\end{gathered}
\end{equation}

\noindent Here, the population of the $i$-th state $f_i$ can be determined as

\begin{equation} \label{eq7}
\begin{gathered}
\textit{f}_i=\frac{n_i}{\textit{g}_i}
\end{gathered}
\end{equation}

\noindent and  (\ref{eq6}), via the Stirling approximation $\ln N!\approx N\ln N$ valid for large $N$, gives a simplified expression for the entropy (\ref{eq5}):

\begin{equation} \label{eq8}
\begin{gathered}
S=k_{\text{B}}\sum\limits_i\textit{g}_i\left[ (1+f_i)\ln(1+f_i)-f_i\ln f_i \right].
\end{gathered}
\end{equation}

The equilibrium state of a complete system of photons corresponds to such a distribution of $n_i$ and~$f_i$ that provides maximum entropy. Additionally, it is necessary to ensure the photon system equilibrium with certain external environment (thermostat) at a temperature $T$ and a constant volume $V$. As an additional condition, we require the conservation of the total energy $U$ of the system:

\begin{equation} \label{eq9}
\begin{gathered}
U=k_{\text{B}}\sum\limits_i E_in_i=\sum\limits_i E_i\textit{g}_if_i=\text{const}\,,
\end{gathered}
\end{equation}

\noindent where $E_i$ is the energy of the $i$-th state.

Following the usual procedure \cite{Goldin}, the maximum of function (\ref{eq5}) upon the condition (\ref{eq9}) is realized when the following equation holds:

\begin{equation} \label{eq10}
\begin{gathered}
\rd S-\lambda \rd U=0,
\end{gathered}
\end{equation}

\noindent where $\lambda$ is the Lagrange multiplier and $\rd S$, $\rd U$ symbolize the full differentials with respect to variables~$f_i$. From equations (\ref{eq8}) and (\ref{eq9}) one easily obtains

\begin{equation} \label{eq11}
\begin{gathered}
\rd S=k_{\text{B}}\sum\limits_i\textit{g}_i\frac{\rd}{\rd f_i}\left[ (1+f_i)\ln(1+f_i)-f_i\ln f_i \right]\rd f_i,
\end{gathered}
\end{equation}

\begin{equation} \label{eq12}
\begin{gathered}
\rd U=\sum\limits_i E_i\textit{g}_i\rd f_i
\end{gathered}
\end{equation}
 and since equation (\ref{eq10}) must hold for any $\rd f_i$, it reduces to the set of identical equations for every $i$-th summand of the sums in (\ref{eq8}) and (\ref{eq9}) so that the subscript ``$i$'' can be omitted:

\begin{equation} \label{eq13}
\begin{gathered}
\frac{\rd}{\rd f}\left[ (1+f)\ln(1+f)-f\ln f \right]-\frac{\lambda}{k_{\text{B}}}E=0.
\end{gathered}
\end{equation}
 This result leads to the usual Bose-Einstein distribution and discloses the meaning of the Lagrange multiplier, $\lambda=T^{-1}$, where $T$ is the absolute temperature:

\begin{equation} \label{eq14}
\begin{gathered}
f(E)=\frac{1}{\exp(\lambda\frac{E}{k_{\text{B}}})-1}=\frac{1}{\exp(\frac{E}{k_{\text{B}}T})-1}.
\end{gathered}
\end{equation}

The energy distribution of the photon concentration ${\rd N}/{\rd E}$ can be found from (\ref{eq14}) as

\begin{equation} \label{eq15}
\begin{gathered}
\frac{\rd N}{\rd E}=\textit{g}(E)f(E),
\end{gathered}
\end{equation}

\noindent where $\textit{g}(E)$ is the density of photon states \cite{Goldin} so that the full number of states in the energy interval $(E, E+\rd E)$ per unit volume is

\begin{equation} \label{eq16}
\begin{gathered}
\textit{g}(E)\rd E=\frac{8\piup}{(c{h})^3}E^2\rd E,
\end{gathered}
\end{equation}

\noindent where ${h}$ is the Planck's constant, and $c$ is the speed of light in vacuum.

The results (\ref{eq14})--(\ref{eq16})  are obtained for the ``classic'' thermodynamic systems where the number of particles and available states tend to infinity. According to the last paragraph of section~\ref{sec1}, in this paper we deal with the systems of mesoscopic scales. Thus, a further employment of the above approach needs an additional justification. First of all, the assumptions $g_i \gg 1$, $n_i \gg 1$ , necessary for transition from equation (\ref{eq6}) to (\ref{eq8}), should be verified. For example, according to (\ref{eq16}), for the radiation wavelength $\lambda=0.5$\textmu{}m (the photon energy $E={h}c/\lambda=2*10^{-19}\text{ J}$), the density of states in the energy interval ($E,E+\Delta E$) is $\textit{g}(E)\Delta E=5.05*10^{38}\Delta E\text{ m}^{-3}$, which in a reasonable case $\Delta E=0.01E=4*10^{-21}\text{ J}$ means $\textit{g}(E)\Delta E=2*10^{18}\text{ m}^{-3}$. Accordingly, even in the microscopic volume $\sim10^{2}\mu \text{m}^{3}=10^{-16}$~\textmu{}m$^{3}$ there will be $\sim200$ photon states, which enables the application of the Stirling formula with a sufficient accuracy.

\section{Non-extensive spectral density of the photon’s number} \label{sec3}

The operation described by equations (\ref{eq8})--(\ref{eq16}) and its results are well known but now we reproduce it in application to the NE entropy (\ref{eq4}). To this end, we represent equation (\ref{eq4}) in the form

\begin{equation} \label{eq17}
\begin{gathered}
\frac{S_{\text{NE}}}{k_{\text{B}}}=\frac{W^{1-q}}{1-q}-\frac{1}{1-q}=\frac{1}{1-q}\re^{(1-q)S/k_{\text{B}}}-\frac{1}{1-q}
\end{gathered}
\end{equation}

\noindent and substitute $S_{\text{NE}}$ instead of $S$ into (\ref{eq10}). According to (\ref{eq17}), $\rd S_{\text{NE}}=\re^{(1-q)S/k_{\text{B}}}\rd S$, and then the analog of equation (\ref{eq10}) appears in the form

\begin{equation} \label{eq18}
\begin{gathered}
\rd S_{\text{NE}}-\lambda \rd U=\re^{(1-q)S/k_{\text{B}}}\rd S-\lambda \rd U=0.
\end{gathered}
\end{equation}

Obviously, its solution can be performed similarly to the chain (\ref{eq11}) – (\ref{eq14}) with the only correction in the Lagrange multiplier which in case of equation (\ref{eq18}) is equal  to $\lambda = \left[ T\re^{(1-q)S/k_{\text{B}}}\right]^{-1}$. Finally, we find that in case of NE entropy (\ref{eq17}), (\ref{eq18}), the equilibrium distribution of photons is described by equation

\begin{equation} \label{eq19}
\begin{gathered}
f_{\text{NE}}(E)=\frac{1}{\exp(\frac{E\Delta}{k_{\text{B}}T})-1},
\end{gathered}
\end{equation}

\noindent where

\begin{equation} \label{eq20}
\begin{gathered}
\Delta=\exp\left[(q-1)\frac{S}{k_{\text{B}}}\right].
\end{gathered}
\end{equation}

\noindent Notably, the right-hand side of equation (\ref{eq19}) also depends on the population function [see (\ref{eq8})], and~(\ref{eq19}) provides no explicit expression for $f_{\text{NE}}(E)$  but, in the first approximation, one may suppose that $S$ in (\ref{eq20}) is determined as the ``usual'' extensive entropy of the black-body radiation, $S=\frac{4}{3}bVT^3$, where $b=(8\piup^5k_{\text{B}}^4)/[15({h}c)^3]$ \cite{Kelly} and $V$ is the physical volume of the non-extensive thermodynamic system under consideration. The energy (spectral) distribution of photons upon the NE conditions follows immediately from (\ref{eq15}), (\ref{eq16}) and (\ref{eq19}):

\begin{equation} \label{eq21}
\begin{gathered}
\frac{\rd N}{\rd E}=
\textit{g}(E)f_{\text{NE}}(E)=\frac{8\piup E^2}{(c{h})^3}\frac{1}{\exp({E\Delta}/{k_{\text{B}}T})-1}.
\end{gathered}
\end{equation}

Equation (\ref{eq21}) is the main result of this work describing the photon density distribution over the energies, depending on the non-extensiveness parameter $q$; the parameter itself should be determined from additional reasoning taking the system nature and properties into account. In the current approximation, while $\Delta$ does not depend on the energy, the result (\ref{eq21}) can be treated as the classical Planck distribution (\ref{eq14})--(\ref{eq16}) provided that $T$ is replaced by the effective temperature $T_{\text{eff}}=T/\Delta > T$. This means that, due to non-extensive properties, the system irradiates as if it were a ``usual'' extensive system but with higher temperature ($T_{\text{eff}}>T$)  because the exponent in (\ref{eq20}) is negative, and $\Delta < 1$.

\section{Results and discussion} \label{sec4}

The possibility to characterize the non-additivity effects by introducing the effective temperature $T_{\text{eff}}$ seems to be the most spectacular result of the above section. This fact has its analogs in some earlier considerations based on the non-extensive entropy analysis. The possibility to preserve the usual thermodynamic relations by introducing the ``physical'', or ``renormalized'' temperature \cite{Abe,Martinez,Tsallis1998} is an attractive feature of the Tsallis approach but it is linked with some simplifying suggestions. In our case, the supposition that $\Delta$ in (\ref{eq20}), (\ref{eq21}) does not depend on energy is crucial but needs additional substantiation because $\Delta$ is determined by the ``additive'' entropy $S$ which does depend on the energy. To elucidate the situation, some general considerations on the nature and origin of the non-additive effects should be also mentioned. In particular, in the Tsallis statistics, the parameter $q$ can be considered within the framework of fractal (multifractal) structures \cite{Deppman}. The thermodynamics of fractal structures is based on the consideration of two ratios: 
\begin{enumerate}
\item The ratio $\langle E \rangle/\langle k_{\text{B}}T \rangle$ of the average internal and kinetic energies of the system. This relationship is constant for the entire subset of systems. 

\item The ratio $E/(k_{\text{B}}T)$ of the local (random) values of energy, which changes in accordance with the distribution function of subsets of the thermodynamic system over energy levels. 
	\end{enumerate}

The difference between these ratios should be taken into account upon the ``effective temperature'' interpretation. In equation (\ref{eq21}), just a random value of energy appears whereas the average entropy $S$ in~(\ref{eq20}) does not depend on the random energy $E$ and, therefore, $\Delta$ can be, to a certain degree, considered as the energy-independent quantity.

Due to this approximation, the spectral density expression (\ref{eq21}) appears in the compact and easily interpretable form, which is advantageous compared with the known result \cite{Tsallis1995}, also derived from the Tsallis entropy. However, the genuine area of its validity is not clear and requires additional investigations, first of all in application to real experimentally observable systems, which can be performed via a consistent numerical examination.

In this process, the so far undetermined phenomenological parameters of the non-extensive description can be reasonably defined. According to equation (\ref{eq20}), the decisive role is played by the product $(1-q)V$, that is, the non-extensiveness parameter $q$ should be determined in conjunction with the volume $V$ whose value is dictated by the spatial size of the system of interacting emitters. In the case of combustion plasma, we are interested in \cite{Zolotko, Vishnyakov, Dragan}, this size is not well known a priori but it is of the order of several to tens of micrometers.

Additionally, the effects of long-range correlations also contribute to the system non-extensiveness. They manifest themselves in the observable behavior of the equilibrium radiation which exhibits a transition from logarithmic asymptotics (Boltzmann entropy) to power asymptotics (Tsallis entropy) with the growing frequency (photon energy). A direct calculation of the degree of non-extensiveness for thermodynamic systems is associated with great difficulties in determining the large thermodynamic potential \cite{Deppman}. However, in the first approximation, it is reasonable to assume that the parameter $q$ preserves a constant value over the whole range of the photon energies. This assumption is consistent with experimental data on the energy distribution of the continuous spectrum in the radiation of a metallized flame, which were used to determine the color temperature \cite{Dragan}.

Now, our task is the numerical study of the thermodynamic characteristics with different sets of phenomenological parameters to obtain the dataset suitable for comparison with experiment, which, ultimately, will enable us to extract these parameters via the procedure of best fitting between the experimental and numerical data. As the first step, and to illustrate the effect of the entropy non-additivity on radiative characteristics, numerical modelling was carried out in the temperature range 2000--3000 K, which is typical of the combustion products of metallized compositions \cite{Vishnyakov}. Another important parameter of equations (\ref{eq20}) and (\ref{eq21}), the system volume, can be approximately evaluated via the following chain of reasoning.

The energy density of the photon gas \cite{Kelly} is $u=bT^4$; the power irradiated from a unit area of the black-body surface is $\varepsilon=(c/4)u$  where $c$ is the light velocity \cite{Goldin}.  Then, the total power irradiated from a sphere of radius $R$ is equal to

\begin{displaymath} 
\begin{gathered}
W=\varepsilon4\piup R^2=\piup c uR^2.
\end{gathered}
\end{displaymath}

\noindent One may consider this sphere as an effective volume of the radiating system. Once the total power irradiated by the system is known, this equation determines the effective sphere radius and, consequently, its volume:

\begin{displaymath} 
\begin{gathered}
V=\frac{4}{3}\piup R^3=\frac{4}{3}\piup \bigg( \frac{W}{\piup cu} \bigg)^{3/2}=\frac{4}{3 \sqrt{\piup}}\bigg( \frac{W}{cbT^4} \bigg)^{3/2}.
\end{gathered}
\end{displaymath}

\noindent For the metal-oxide agglomerates in combustion plasma, the irradiated power $W$ can be available from experimental data \cite{Zagorodny}; for example, supposing $W = 1$~mW and $T = 3000$~K, we obtain the referential value $W \approx 3\cdot10^{-16}\text{ m}^3=300$~\textmu{}m$^3$.

\begin{figure}[!t]
	\centerline{\includegraphics{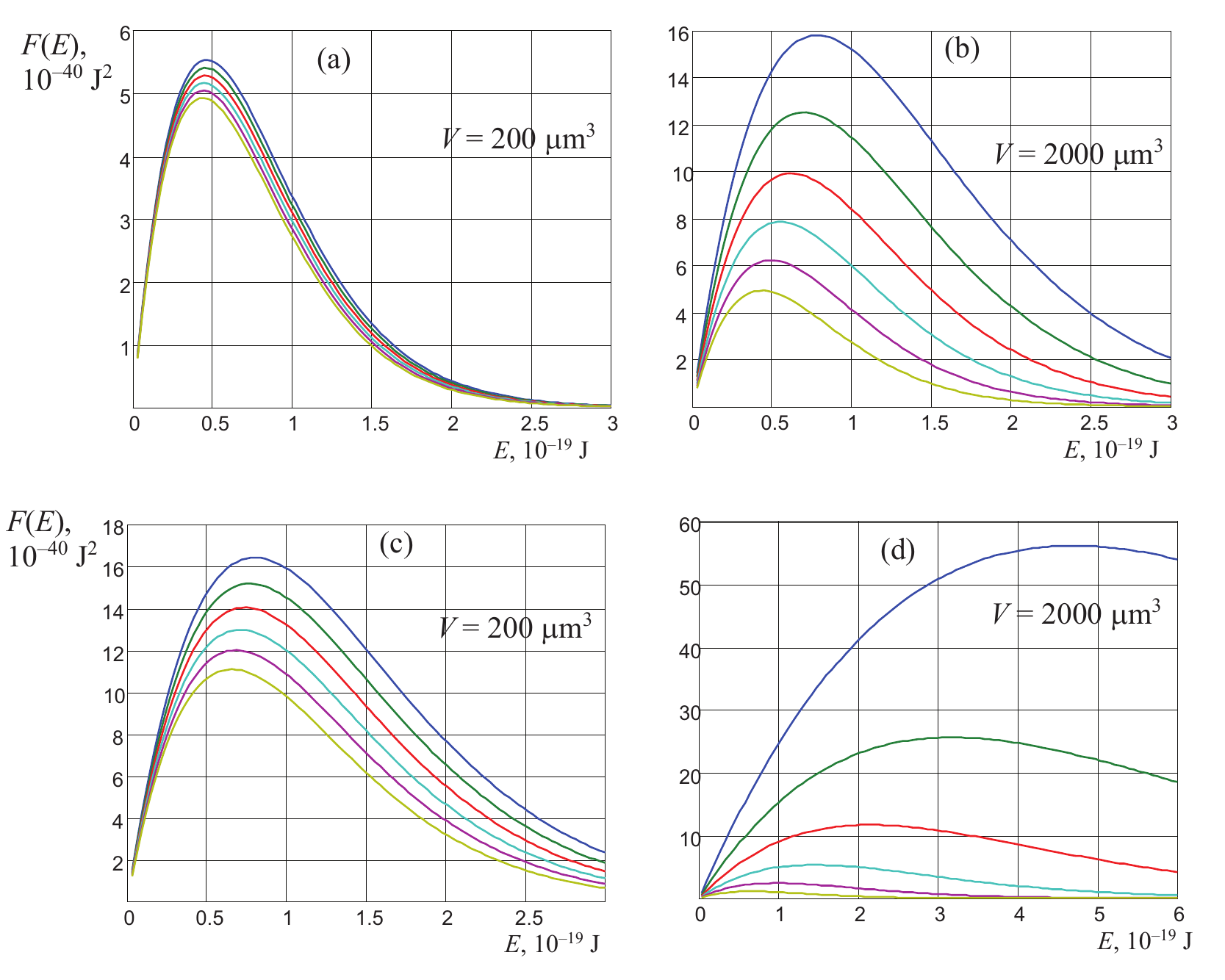}}
	\caption{Normalized spectral distributions of the photon density (\ref{eq22}) for different values of the parameter~$q$ and different system volumes (indicated on the panels): (a, b) for $T=2000$ K and (c, d) $T=3000$~K. (Blue) $q=0.9995$, (green) $q=0.9996$, (red) $q=0.9997$, (cyan) $q=0.9998$, (magenta) $q=0.9999$, (yellow) $q=1.0$. Yellow curves coincide with the classical Planck distribution of (\ref{eq14}), (\ref{eq15}).}
	\label{figure1}
\end{figure}

Of course, in real situations the value of $V$ can differ from this referential value. In this paper, for the illustrative model calculations we accept the two values of the system volume, $V=200$~\textmu{}m$^3$ and $V=2000$~\textmu{}$m^3$ (see figure~\ref{figure1}). The values of the parameter $q$ are chosen in the range $0.9995 \leqslant q \leqslant 1$. For convenience and simplicity of presentation, we employ the photon density distribution (\ref{eq21}) in the following normalized form:

\begin{equation} \label{eq22}
\begin{gathered}
F(E)=\frac{c^3{h}^3}{8\piup}\frac{\rd N}{\rd E}=\frac{E^2}{\exp({E\Delta}/{k_{\text{B}}T})-1}.
\end{gathered}
\end{equation}

\noindent The calculations are performed for $T=2000$ K and $T=3000$ K and the photon energy range $0 < E < 5$~eV, which are of interest for metallized thermal combustion plasma \cite{Vishnyakov, Dragan}.

The results are shown in figure~\ref{figure1}. As it follows from the graphs, the radiation intensity in the system of non-extensive radiators exceeds the Planck radiation intensity (illustrated by yellow curves) in the whole energy range, which is expected in view of the negative sign of the exponent in (\ref{eq20}) and $\Delta < 1$. The second peculiar feature of a considered non-extensive system is that the spectral density maximum is shifted towards higher photon energies. Comparison of the curves in the left-hand and right-hand column of figure~\ref{figure1} apparently testifies that the influence of the non-extensiveness grows with the system volume, which looks contradictory because the non-additive interactions are more influential in small systems \cite{Hill}. This seeming controversy is explained by the fact that, in our calculations, the parameters $q$ and $V$ are formally considered as independent quantities whereas actually $q$ depends on the volume and drastically decreases with growing $V$ \cite{Tsallis, Kolesnichenko,Tsallis1995} (note, in addition, that the huge difference in pertinent volumes prevents the direct juxtaposition of our results with those of \cite{Tsallis1995}, adapted to cosmic radiation). Additionally, the curves of figure~\ref{figure1} show that in the high-frequency region of the spectrum, the radiation behavior deviates progressively from the Planck’s law, in qualitative agreement with the known observations \cite{Zolotko, Vishnyakov, Dragan}.

Note that the form (\ref{eq22}) is suitable for the theoretical analysis but in practical situations the photon density is rather exotic; however, the knowledge of the function $F(E)$  gives access to the more familiar frequency- and wavelength-dependent radiation energy density distributions \cite{Balescu, Goldin, Kubo} per unit frequency interval:

\begin{displaymath} 
\begin{gathered}
u_\nu(\nu)=\frac{8\piup {h}\nu^3}{c^3}\frac{1}{\exp(\frac{{h}\nu}{k_{\text{B}}T}\Delta)-1}=\frac{8\piup}{c^3{h}^2} \left[ {h}\nu F({h}\nu) \right] 
\end{gathered}
\end{displaymath}

\noindent and per unit wavelength interval:

\begin{displaymath} 
\begin{gathered}
u_\lambda(\lambda)=\frac{8\piup {h}c}{\lambda^5}\frac{1}{\exp(\frac{{h}c}{\lambda k_{\text{B}}T}\Delta)-1}=\frac{8\piup}{({h}c\lambda)^2} \left[ \frac{{h}c}{\lambda} F\Big(\frac{{h}c}{\lambda}\Big) \right]. 
\end{gathered}
\end{displaymath}

\noindent With these formulae, all the characteristic features of the curves $F(E)$  discussed in figure~\ref{figure1} can be extended, with obvious modifications, to the dependences $u_\nu(\nu)$ and $u_\lambda(\lambda)$, immediately observable in experiments. 

\begin{figure}[!t]
	\centerline{\includegraphics{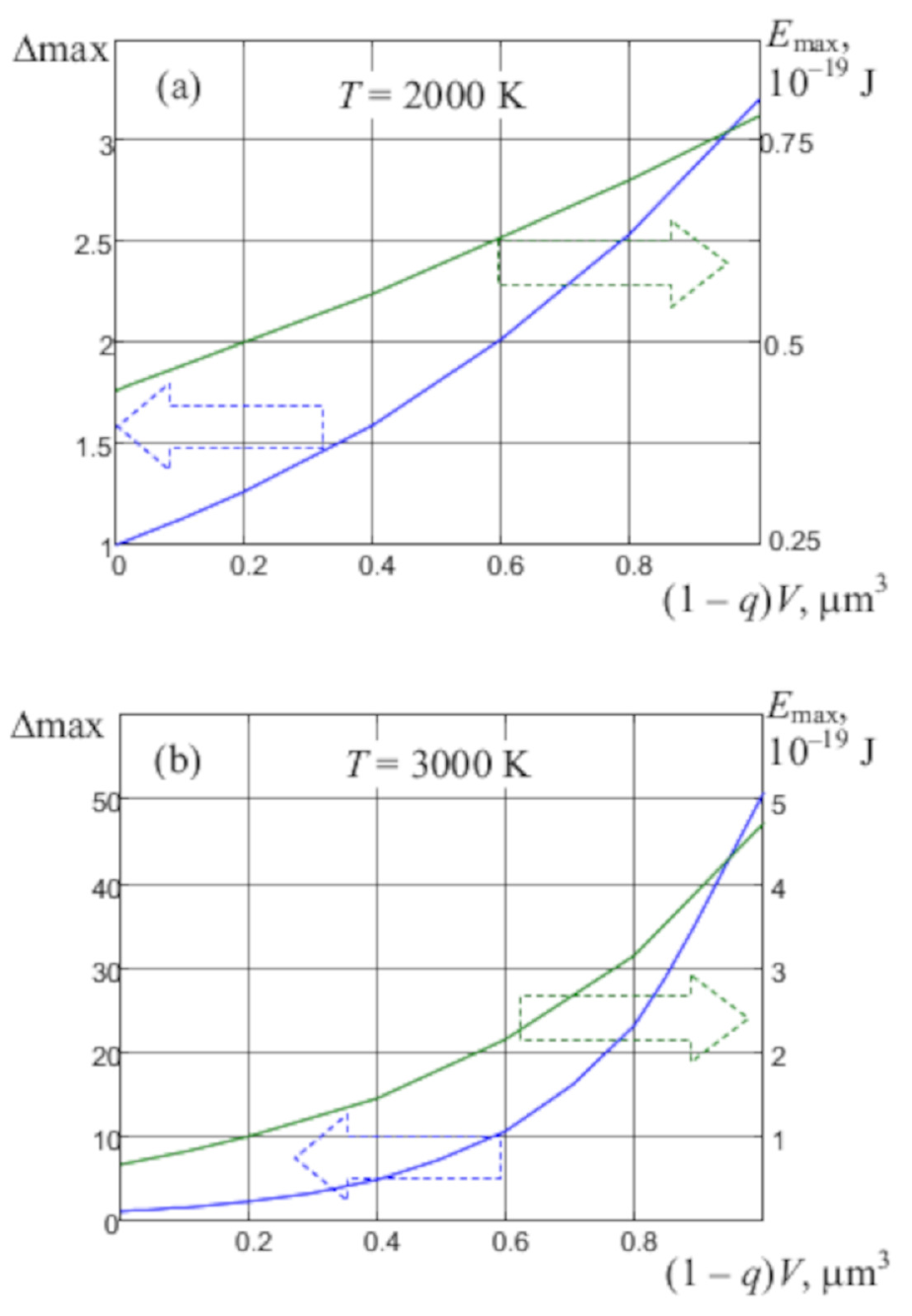}}
	\caption{Relative maximum (\ref{eq23}) of the photon density distribution (\ref{eq22}) (blue curves, left-hand scales) and the corresponding photon energy $E_{\mathrm{max}}$ (green curves, right-hand scales) vs the non-extensiveness parameter $(1-q)V$ for (a) $T=2000$ K and (b) $T=3000$ K.}
	\label{figure2}
\end{figure}

The main quantitative characteristics of the spectral distribution (the energy corresponding to the maximum spectral density $E_{\mathrm{max}}$ and the spectral density maximum magnitude) are illustrated more in  detail in figure~\ref{figure2}. The maximum absolute value is presented in relative units (with respect to the maximum spectral density of the classical Planck distribution at the same temperature):

\begin{equation} \label{eq23}
\begin{gathered}
\Delta \mathrm{max}=\frac{\mathrm{max}[F(E)]}{\mathrm{max}[F(E)]_{q=1}}.
\end{gathered}
\end{equation}

\noindent Due to the non-extensiveness, the maximum absolute value grows, sometimes rather remarkably (see figure~\ref{figure1}d and the blue curve in figure~\ref{figure2}b). The spectral maximum shift due to non-extensiveness behaves almost linearly with an increase of the parameter $(1-q)V$  at a lower temperature but shows a super-linear growth at higher temperatures (see the green curves in figures~\ref{figure2}a, ~\ref{figure2}b).

Note that the extremely high values of $\Delta \text{max}$ on figure~\ref{figure2}b (up to 50 at $(1-q)V=1$~\textmu{}m$^3$ are mostly illustrative; in practice, such high values of the product $(1-q)V$ are non-physical because with the growing volume, $q$ tends to  unity so that the whole product satisfies the inequality $(1-q)V \ll 1$~\textmu{}m$^3$ in most real situations.

\section{Conclusion}

The problem of  deviation of the thermal radiation spectrum of various systems (from cosmic clouds and galaxies in astrophysics to the non-equilibrium fuel plasma in combustion techniques) from the Planck law has long been discussed in the literature. It is especially important for determining the temperature of condensed substances by the ``color'' method based on comparing the observed radiation spectrum with the Planck one \cite{Vishnyakov, Dragan}. In particular, the approach based on the non-extensive thermodynamics was realized in application to the cosmic radiation \cite{Tsallis1995}. At the same time, to the best of the authors’ knowledge, there was no detailed discussion and attempts to employ the non-extensive thermodynamics in systems of micro-radiators characteristic of the combustion plasma.

This paper has demonstrated that the application of Tsallis thermodynamics makes it possible to take into account the influence of the system entropy non-extensiveness on its radiative characteristics. The model calculations are performed for the micrometer-size interacting plasma clusters that are in a radiative thermal contact with a thermostat of the temperature range $2000-3000$ K typical of chemical combustion. Calculations show the influence of the Tsallis non-extensiveness parameter $q$  on the spectral distribution of the radiation energy and the spectral position of the radiation maximum. Specifically, it is shown that the overall radiation energy grows, and the maximum position moves almost linearly towards the higher energies  with an increasing difference between the parameter $q$ and 1.

On the other hand, our calculations confirm the fact that the non-Planck features of the radiation spectra of the metallized fuel combustion products that were revealed while determining the temperature by the color method \cite{Dragan}, indicate the entropy non-extensiveness for the emitter subsystems. It is suggested that the violation of the classical Planck’s law observed in the combustion plasma radiation can be caused by interactions between submicrometer plasma particles, clusters, agglomerates that occur in the chemical reaction zone during the combustion, for example, of metal particles and charged condensed particles. Generally, one can expect that the entropy non-extensiveness will noticeably manifest itself in the similar nonequilibrium systems, especially in chemically reacting ones.

In particular, the earlier experimental data relating the emission spectra of burning metallized compositions \cite{Zolotko, Vishnyakov, Dragan} should be revisited with an explicit employment of the non-additive thermodynamics concepts. In our opinion, the systems of chemically reacting condensed micro- and nanoparticles will become promising objects for application of the non-extensive thermodynamics; especially, the studies of their emission spectra will clarify the situation regarding the physical meaning and the value of the Tsallis’s phenomenological parameter $q$. For example, the shift of the emission maximum (see green curves in figure~\ref{figure2}) can be used for  experimental determination of the parameter $q$, disclosing its physical nature and its relations with physical properties of the radiating system.


\bibliographystyle{cmpj}

\begin{thebibliography}{10}
	\bibitem{Zolotko} Zolotko~A.~N., Vovchuk~Ya.~I., Poletaev~N.~I.,  Florko~A.~V., Fiz. Goreniya Vzryva, 1996, \textbf{32}, 24--33, (in Russian).
	\bibitem{Vishnyakov} 
	Vishnyakov V. I., Dragan G. S., Phys. Rev. E, 2006, \textbf{73}, No. 2, 026403 (7 pages),\\\doi{10.1103/PhysRevE.73.026403}.
	\bibitem{Lautenberger} 
	Lautenberger C., Tien C.L., Lee K.Y., Stretton A.J., In: SFPE Handbook of Fire Protection Engineering, Hurley M.~J. et al. (Eds.). Springer, New York, 2016, 102--137, \doi{10.1007/978-1-4939-2565-0_4}.
	\bibitem{Ning} Ning~J., Wang~W., Ning~X., Sievers~D.~E., Garmestani~H.,  Liang~S.~Y., Materials, 2020, \textbf{13}, 8, \\\doi{10.3390/ma13081988}.
	\bibitem{Lubin} Lubin~P., Villela~T., Epstein~G., Smoot~G., Astrophys. J., 1985, \textbf{298}, L1-L5, \doi{10.1086/184555}.
	\bibitem{Trigger} Trigger~S.~A.,  Phys. Lett. A, 2007, \textbf{370}, 365--369, \doi{10.1016/j.physleta.2007.05.084}.
	\bibitem{Planck} Planck~M., Ann. Phys., 1901, \textbf{4}, 553--563, \doi{10.1002/andp.19013090310}.
	\bibitem{Zagorodny} Zagorodny~A.~G., Trigger~S.~A., Bull. Lebedev Phys. Inst., 2018, \textbf{45}, 159--164, \\\doi{10.3103/S106833561805007X}.
	\bibitem{Dragan} Dragan~G.~S., J. Exp. Theor. Phys., 2004, \textbf{98}, 3, 503--507, \doi{10.1134/1.1705702}.
	\bibitem{Tsallis} Tsallis~C., J. Stat. Phys., 1988, \textbf{52}, No.~1--2, 479--487, \doi{10.1007/BF01016429}.
	\bibitem{Kolesnichenko} Kolesnichenko~A.~V., Preprints of Keldysh Institute, 2020, \textbf{35}, 1--28, \doi{10.20948/prepr-2020-35}, (in Russian).
	\bibitem{Hill} Hill~T.~L., Thermodynamics of Small Systems, Parts I \& II, Dover Publications, New York, 2013.
	\bibitem{Balescu} Balescu~R., Statistical Mechanics of Charged Particles, London, Interscience Publ., 1963.
	\bibitem{Pavlos} Pavlos~G.~P., Iliopoulos~A.~C., Tsoutsouras~V.~G., Sarafopoulos~D.~V., Sfiris~D.~S., Karakatsanis~L.~P., \\Pavlos~E.~G.,  Physica A, 2011, \textbf{390}, No.~15, 2819--2839, \doi{10.1016/j.physa.2011.03.005}.
	\bibitem{Yoon2019} Yoon~P.~H.,  Entropy, 2019, \textbf{21}, No.~9, 820, \doi{10.3390/e21090820}. 
	\bibitem{Yoon2020} Yoon~P.~H.,  Eur. Phys. J. Spec. Top., 2020, \textbf{229}, No.~5, 819--840, \doi{10.1140/epjst/e2020-900215-4}. 
	\bibitem{Strzalka} Strza\l{}ka~D., Grabowski~F., Fundam. Inform., 2008, \textbf{85}, No.~1--4, 455--464.
    \bibitem{Ghanbari} Ghanbari~A., Khordad~R., Ghaderi-Zefrehei~M.,  Physica B, 2021, \textbf{624}, 413448, \\\doi{10.1016/j.physb.2021.413448}.
    \bibitem{Groot} De Groot~S.~R., Mazur~P., Non-equilibrium Thermodynamics. Dover Publications, New York, 1984.
    \bibitem{Olemskoi} Olemskoi~A.~I., Synergetics of Complex Systems: Phenomenology and Statistical Theory, Krasand, \\Moscow, 2009.
    \bibitem{Goldin} Goldin~L.~L., Novikova~G.~I., Vvedenie v Atomnuju Fiziku, Nauka, \\Moscow, 1988, (in Russian).
    \bibitem{Kelly} Kelly~R.~E., Am. J. Phys., 1981, \textbf{49}, 714--719, \doi{10.1119/1.12416}.
    \bibitem{Abe} Abe~S., Martinez~S., Pennini~F., Plastino~A.,  Phys. Lett. A, 2001, \textbf{281}, No.~2--3, 126--130, \\\doi{10.1016/S0375-9601(01)00127-X}.
    \bibitem{Martinez} Martinez~S., Pennini~F., Plastino~A., Tessone~C.,  Physica A, 2001, \textbf{295}, No.~1--2, 224--229, \\\doi{10.1016/S0378-4371(01)00078-4}.
    \bibitem{Tsallis1998} Tsallis~C., Mendes~R., Plastino~A.~R., Physica A, 1998, \textbf{261}, No.~3--4, 534--554, \\\doi{10.1016/S0378-4371(98)00437-3}.
    \bibitem{Deppman} Deppman~A., Phys. Rev. D, 2016, \textbf{93}, 054001, \doi{10.1103/PhysRevD.93.054001}.
    \bibitem{Tsallis1995} Tsallis~C., Barreto~F.~S., Loh~E.~D.,  Phys. Rev. E, 1995, \textbf{52}, No.~2, 1447--1451, \doi{10.1103/PhysRevE.52.1447}.
    \bibitem{Kubo} Kubo~R., Statistical Mechanics, North-Holland, New York, 1988.

\end{thebibliography}

%
%

\ukrainianpart

\title{Неекстенсивна термодинамiка випромiнювання в гетерогенній термічній плазмі}
\author{Г. С. Драган, В. В. Кутаров, О. Я. Бекшаев}
\address{Науково-дослідний інститут фізики, Одеський національний університет імені І.~І. Мечникова, Дворянська, 2, Одеса, Україна, 65082}

%
%
%

\makeukrtitle

\begin{abstract}
\tolerance=3000%
Термодинамічні характеристики випромінювання конденсованих продуктів згоряння у вигляді агломератів, утворених наночастинками оксидів металів, демонструють відхилення від класичного закону Планка. Пропонується інтерпретувати ці відхилення з точки зору неадитивної ентропії фотонної системи, яка взаємодіє з гетерогенними продуктами горіння, що дозволяє використати для їх опису неекстенсивну термодинаміку Цалліса. При цьому вважається, що неадитивний характер ентропії випромінювання в гетерогенній плазмі можна пояснити впливом далекодіючих взаємодій та нерівноважними фізико-хімічними процесами. Отримано вираз для густини розподілу фотонів за енергіями, що базується на застосуванні феноменологічного параметра неекстенсивності $q$, який у першому наближенні не залежить від енергії. У цьому випадку ''неекстенсивний'' закон Планка можна звести до ``звичайного'' розподілу Планка шляхом введення ``ефективної температури'', яка перевищує реальну температуру. Чисельне моделювання показало, що спектральний розподіл фотонів, положення та величина його максимуму залежать від значення параметра $q$, що може бути використано для експериментального визначення цього параметру та виявлення його фізичної природи і походження.
\keywords ентропія, випромінювання, гетерогенна плазма, термодинаміка Цалліса

\end{abstract}

\end{document}